\documentclass{iopart}
\usepackage{graphicx}
\usepackage{iopams}  
\begin{document}

\title[Phase diagrams of triangular and kagome antiferromagnets]
{Magnetic phase diagrams of classical triangular and kagome antiferromagnets} 

\author{M V Gvozdikova,$^1$ P-E Melchy$^2$ and M E Zhitomirsky$^{2,3}$}

\address{$^1$ Department of Physics, Kharkov National University,
61077 Kharkov, Ukraine}

\address{$^2$ Service de Physique Statistique, Magn\'etisme et Supraconductivit\'e,
UMR-E9001 CEA-INAC/UJF, 17 rue des Martyrs, 38054 Grenoble, France} 

\address{$^3$ Max-Planck-Institut f\"ur Physik Komplexer Systeme, N\"othnitzer str.\
38,  D-01187 Dresden, Germany 
}
\ead{mike.zhitomirsky@cea.fr}
\begin{abstract}
We investigate the effect of geometrical frustration on the $H$--$T$ phase diagrams of the
classical Heisenberg antiferromagnets on triangular and  kagome lattices.
The phase diagrams for the two models are obtained from large scale Monte Carlo 
simulations. For the kagome antiferromagnet thermal fluctuations are unable 
to lift completely degeneracy and stabilize translationally disordered
multipolar phases. We find a substantial difference 
in the temperature scales of the order by disorder effect related
to different degeneracy of the low- and the high-field classical ground states
in the kagome antiferromagnet.
In the low-field regime, the Kosterlitz-Thouless transition into
a spin-nematic phase is produced by unbinding of half-quantum vortices. 
\end{abstract}

\submitto{\JPCM}


\section{Introduction}

One of the important issues in the field of highly frustrated magnets 
is the universality of the order by disorder phenomenon \cite{Villain80,Shender82}.
If the `accidental' degeneracy of a frustrated spin system is described by
a few continuous  degrees of freedom  
\cite{Shender82,Kawamura84,Henley89,Chandra90,Chubukov91,Weber03}
or by an infinite but non-extensive number of such parameters 
\cite{Gvozdikova05,Bergman07}
the order by disorder mechanism is known to stabilize a magnetically ordered state
at low temperatures. In the case of a macroscopic number of zero-energy modes
the outcome is less universal and only a partial
lifting of the degeneracy, if any at all, may take place \cite{Moessner98,Tchernyshyov03,Henley06}.

In this work we study numerically the entropic order by disorder selection produced by thermal
fluctuations in two classical antiferromagnets on a triangular 
and a kagome lattice in external magnetic field. The external field changes continuously 
the classical ground-states manifold of a frustrated magnet providing 
an experimental tool to control and  to modify the effect of fluctuations
\cite{Zhitomirsky00,Moessner09}. The magnetic phase diagrams for the two models
have been discussed qualitatively in the earlier works \cite{Kawamura85,Zhitomirsky02}.
Here we present results of large-scale Monte Carlo simulations, which allow precise
mapping of the transition boundaries and clarify nature of the low-temperature
phases. In addition to providing new insights into the order
by disorder effect in classical frustrated magnets, our results open a possibility 
of quantitative comparison with the experimental phase diagrams of recently discovered 
triangular antiferromagnets $\rm RbFe(MoO_4)_2$ \cite{Svistov03,Svistov06} and 
$\rm Rb_4Mn(MoO_4)_3$ \cite{Ishii09}.
 
\begin{figure}[t]
\centerline{
\includegraphics[width=0.75\columnwidth]{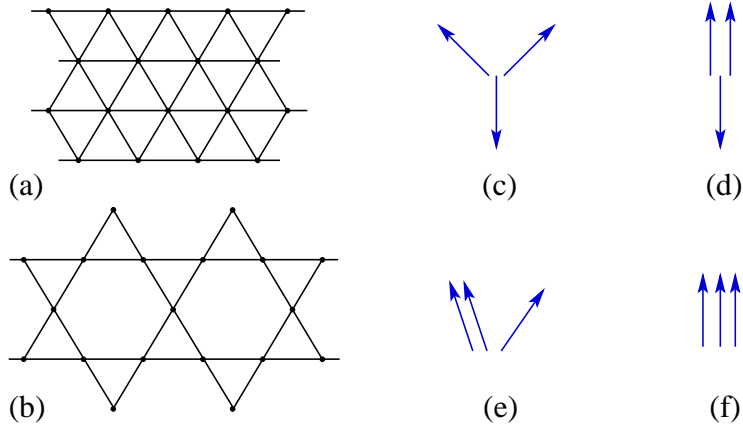}
}
\caption{Triangular (a) and kagome (b) lattices.
Equilibrium three-sublattice configurations 
in different field regimes: (c) for $0<H<\frac{1}{3}H_s$, 
(d) for $H=\frac{1}{3}H_s$, (e) for $\frac{1}{3}H_s<H<H_s$ and (f) for $H>H_s$.
}
\label{fig1}
\end{figure}

The nearest-neighbor exchange Hamiltonian with a Zeeman term is given by
\begin{equation}
{\cal H} = J \sum_{\langle ij\rangle} {\bf S}_i\cdot {\bf S}_j - 
H\sum_i S^z_i \ ,
\label{H}
\end{equation}
where ${\bf S}_i$ are three-component classical unit vectors. 
The model (\ref{H}) with an antiferromagnetic exchange $J>0$
is studied for a triangular and  a kagome lattice, see figures~\ref{fig1}(a) and (b).
The common building block of the two lattices, a triangular plaquette,
determines frustration at the shortest length-scale. In zero field 
the antiferromagnetic  bonds on a single spin-triangle cannot be 
simultaneously satisfied and magnetic moments form a $120^\circ$ spin-structure. 
For the triangular antiferromagnet this leads to an ordered magnetic
state with the ordering wavevector ${\bf Q} =(4\pi/3,0)$ and a trivial six-fold 
degeneracy related to permutations of the spin triad.
In contrast, the kagome-lattice model remains infinitely degenerate. 
The coplanar ground-states of the  kagome antiferromagnet can be mapped onto 
different colouring patterns of an exactly solvable three-colour problem
\cite{Chalker92,Huse92}. The number of colourings
grows exponentially with the number of sites $N$ as
\begin{equation}
W \approx 1.1347^N \ .
\label{N3col}
\end{equation} 

In a finite magnetic field the classical ground states may be derived
by rewriting the energy (\ref{H}) as a sum over triangles
\cite{Zhitomirsky00, Lee84}. The classical energy reaches the minimal value for 
the following magnetization of every plaquette:
\begin{equation}
\sum_{i\in\triangle} {\bf S}_i  = 3\frac{H}{H_s} \hat{\bf z}\ ,
\label{constraint}
\end{equation}
where $H_s$ stands for the saturation field with
$H_s =9J$ and $6J$ for the triangular and the kagome antiferromagnet,
respectively. The constraint (\ref{constraint}) leaves three 
parameters, instead of one, for the choice of the spin triad ${\bf S}_i$ at $H<H_s$. 
Physically, extra degrees of freedom correspond to an isotropic susceptibility 
of a spin triangle. Thus, an external magnetic field produces new degeneracy in the case of \
the triangular antiferromagnet and enhances the existing degeneracy for the kagome model.

The entropic degeneracy lifting in the classical triangular antiferromagnet was 
first studied by Kawamura, who calculated the free-energy contribution of 
the harmonic excitations \cite{Kawamura84}. 
Thermal fluctuations lift the degeneracy and
stabilize in the interval $0<H<H_s$ one collinear and two coplanar states. 
These are shown in figures~\ref{fig1}(c)--(e) and will be referred to in 
the following as the coplanar Y-state, $0<H<\frac{1}{3}H_s$,
the coplanar $\mathbb{V}$-state, $\frac{1}{3}H_s<H<H_s$,
and the collinear up-up-down ($uud$) state, $H\sim\frac{1}{3}H_s$.

The harmonic fluctuations in the kagome antiferromagnet are dominated by 
zero-energy modes, which contribute most to the free-energy decrease at 
low temperatures \cite{Chalker92}. Their number for different classical 
ground states (\ref{constraint}) can be counted by using purely geometrical 
arguments. As a result, one ends up with the same sequence of three-sublattice 
configurations in magnetic field, figures~\ref{fig1}(c)--(e) \cite{Zhitomirsky02}. 
In addition, the zero-energy modes favour the $\sqrt{3}\times\sqrt{3}$ 
periodic pattern for the  $\mathbb{V}$-state, whereas no such selection
takes place for the Y- and for the $uud$-state. Note, also that two sublattices 
become identical for $uud$ and $\mathbb{V}$ configurations. Consequently, 
the total number of these states is smaller than (\ref{N3col}) and is given 
by the number of dimer coverings of the dual hexagonal lattice 
\cite{Moessner01,Zhitomirsky02}:
\begin{equation}
W \approx 1.1137^N \ .
\label{Ndimer}
\end{equation} 
Lower degeneracy in the high-field region $\frac{1}{3}H_s<H<H_s$
may significantly affect the fate of the order by disorder selection.

In the following we present the Monte Carlo results for
the phase diagrams of the triangular (section~\ref{sec:triangular})
and of the kagome antiferromagnet (section~\ref{sec:MCkagome}).
In section~\ref{sec:multipolar} we also give a brief symmetry analysis of 
various multipolar states in zero and finite magnetic fields. 
The obtained results are summarized in section~\ref{sec:discussion}.

\section{Triangular antiferromagnet}
\label{sec:triangular}

Monte Carlo simulations have been performed 
using the standard Metropolis algorithm in combination
with the microcanonical over-relaxation steps, see \cite{Zhitomirsky08}
for further details.  Periodic boundary conditions were implemented
for $N=L\times L$ site clusters with the linear size $L$ up to 192. 
At every temperature/magnetic field 
we discarded  $5\times 10^4$  Monte Carlo steps (MCS) for
initial relaxation and data were collected during
subsequent $10^6$ MCS. The error bars were estimated
from  10--20 independent runs initialized by 
different random numbers. 
Instantaneous values of the antiferromagnetic order parameter
\begin{equation}
{\bf m}_{\bf Q} = \frac{1}{N} \sum_i {\bf S}_i \,\rme^{-\rmi{\bf Q}{\bf r}_i}
\ , \qquad {\bf Q} = (\case{4}{3}\pi,0)
\label{AF}
\end{equation}
have been used to measure longitudinal and transverse components
of the staggered spin susceptibility and the corresponding Binder cumulants:
\begin{equation}
\chi^\alpha_{\bf Q} = \frac{N}{T} \langle (m^\alpha_{\bf Q})^2\rangle 
\ , \qquad 
U^\alpha_{\bf Q} = \frac{ \langle (m^\alpha_{\bf Q})^4 \rangle}
{\langle (m^\alpha_{\bf Q})^2\rangle^2} \ , 
\end{equation}
where $\alpha = z,\perp$ and $m_{\bf Q}^\perp = (m^{x2}_{\bf Q}+m^{y2}_{\bf Q})^{1/2}$.

\begin{figure}[t]
\centerline{
\includegraphics[width=0.5\columnwidth]{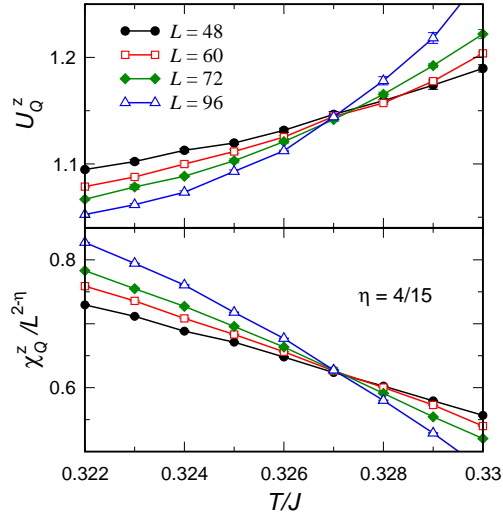}
}
\caption{Transition between the paramagnetic phase and the $uud$-state
in the triangular antiferromagnet at $H/J=1$: the Binder cumulant $U^z_{\bf Q}$
(upper panel) and the staggered susceptibility $\chi^z_{\bf Q}$ (lower panel) 
versus temperature for different lattice sizes $L$.}
\label{fig2}
\end{figure}

Following the standard procedure, the second-order transition may be located 
by the crossing point of the Binder cumulant $U_L(T)$ measured  
for different clusters. 
We  illustrate this method in the top panel of figure~\ref{fig2} 
for the transition between the paramagnetic state
and the $uud$ state  at $H/J=1$. The alternative approach is to use 
the order parameter susceptibility provided the critical exponent $\eta$ 
is known precisely. In the critical region the susceptibility scales as 
\begin{equation}
\label{fss}
\chi = L^{2-\eta} f\bigl(|\tau| L^{1/\nu}\bigr) \ , \quad \tau = 1 - T/T_c \ .
\end{equation}
Hence, the normalized susceptibility $\chi/L^{2-\eta}$ becomes
size-independent at $\tau=0$ and curves for  different $L$ 
plotted as functions of $T$ exhibit a crossing point, similar to 
the behavior of the Binder cumulant. The collinear $uud$ state of the triangular 
antiferromagnet breaks the $\mathbb{Z}_3$ lattice symmetry. The associated phase transition 
belongs to the universality class of the two-dimensional three-state Potts model,
for which the critical exponents are known exactly including $\eta = 4/15$ 
\cite{Chaikin}. 
The bottom panel in figure~\ref{fig2} demonstrates that the normalized susceptibility 
curves for the above value of $\eta$ cross precisely at the same point
as the Binder cumulants. On one side this proves that the critical behavior for 
the transition into the $uud$ state belongs to the three-state Potts universality class, 
on the other side it allows to employ stochastically less noisy $\chi^z_{\bf Q}$ 
to determine the full transition boundary in the $H$--$T$ plane.

At low temperatures the transverse spin components exhibit a quasi-long-range 
order. The Binder cumulant method is not very convenient for a precise location
of the corresponding Kosterlitz-Thouless (KT) transition. 
However, one can still use the normalized transverse susceptibility 
$\chi^\perp_{\bf Q}/L^{2-\eta}$ with the exact KT exponent $\eta = 1/4$ \cite{Chaikin}. 
We also measured the spin stiffness with respect to twists about 
the $\hat{\bf z}$-axis given  for the triangular antiferromagnet by \cite{Lee84} 
\begin{eqnarray}
\rho_s & = &  -\frac{J}{N\sqrt{3}}\, \sum_{\langle ij \rangle}
\, \langle (S_{i}^xS_{j}^x + S_{i}^yS_{j}^y) \rangle 
\label{rhoS} \\
& & \mbox{} - \frac{2J^2}{NT\sqrt{3}} 
\Bigl\langle \Bigl[\,\sum_{\langle ij \rangle} (S_{i}^xS_{j}^y\! -\! 
S_{i}^yS_{j}^x) (\hat{\bf  e}\cdot\boldsymbol{\delta}_{ij}) \Bigr]^2
\Bigr\rangle  \ .
\nonumber
\end{eqnarray}
Here, $\boldsymbol{\delta}_{ij}={\bf r}_{i} - {\bf r}_{j}$ and 
$\hat{\bf e}$ stands for an arbitrary unit vector in the lattice
plane. The spin stiffness shows a universal jump from zero to $\rho_s = 2T_{\rm KT}/\pi$
at the KT-transition. Measurements of $\rho_s(T)$ provide, therefore, an independent 
check for the susceptibility method.

\begin{figure}[t]
\centerline{
\includegraphics[width=0.5\columnwidth]{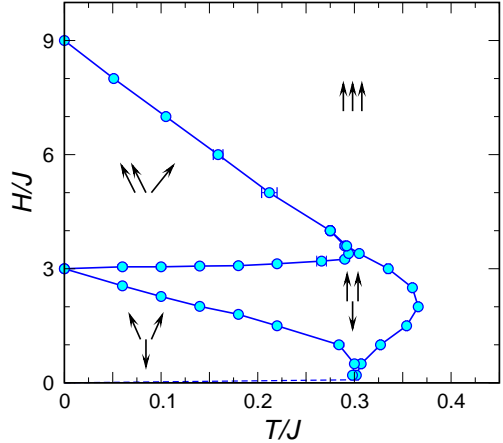}
}
\caption{Magnetic field phase diagram of the Heisenberg triangular-lattice 
antiferromagnet. Transition points determined by the Monte Carlo
simulations are shown by circles. Solid lines are guides for the eye.
Dashed line indicates position of additional low-field transitions.
}
\label{fig3}
\end{figure}

Results of various temperature and magnetic field scans are 
summarized in the $H$--$T$ phase diagram  presented in figure~\ref{fig3}.
The high-temperature paramagnetic phase remains invariant under the
$O(2)\times\mathbb{Z}_3$ symmetry group, where the discrete
symmetry $\mathbb{Z}_3$ is related to the permutation of 
three antiferromagnetic sublattices imposed by lattice translations: 
$\hat{T}_{\bf a}\, {\bf m}_{\bf Q} \rightarrow \rme^{\rmi\bf Qa}\,{\bf m}_{\bf Q}$.
Three ordered phases in figure~\ref{fig3}
break in different way the full symmetry group.
The boundary between the paramagnetic and the $uud$ state is formed by 
the second-order transition points, all other boundaries
correspond to the KT-transitions. The notable feature of 
the phase diagram is a wide region with the collinear 
$uud$ magnetic order \cite{Kawamura85,Lee84}. At $T=0$ this collinear 
spin configuration appears only for one specific value of 
the applied magnetic field $H=\frac{1}{3}H_s$. 
Thermal fluctuations expand its stability range, such that  
the $uud$ state has the highest transition temperature
among the three ordered spin structures.
The collinear ordering produces a clear 1/3-magnetization
plateau, which is most pronounced in the plots
$dM/dH$ versus $H$. 

Overall the phase diagram shown in figure~\ref{fig3} is similar to the diagram 
proposed by Kawamura and Miyashita in \cite{Kawamura85}. The only qualitative 
difference concerns the low-field part. For temperature scans in a constant field 
$H<\frac{1}{3}H_s$ we always find two successive transitions: one from 
the paramagnetic state into the $uud$ phase with the $\mathbb{Z}_3$-order of 
$z$-components  and, then, into 
the Y-state with an algebraic order of the transverse
spin components. There is no direct transition 
between the paramagnetic state and the Y-phase suggested in \cite{Kawamura85}. 
The double transition is completely natural from the symmetry point
of view because ordering processes for longitudinal and transverse spin components 
are generally decoupled  from each other.
Note also that the Y-state and the $uud$-state should be separated
from the disordered state in zero field by additional line(s) of phase 
transitions at $H/J\lesssim 0.1$. The present study remains uncertain about precise
location of these  transition boundaries. Numerical investigation of
the low-field regime requires Monte Carlo simulations of
substantially bigger clusters with $L\sim 10^3$ because of the rapidly growing 
correlation length $\xi$ below $T^*\approx 0.3J$ 
\cite{Kawamura09}.

\section{Kagome antiferromagnet}
\label{sec:kagome}

\subsection{Multipolar phases in zero and finite magnetic fields}
\label{sec:multipolar}

The thermal order by disorder effect in the kagome antiferromagnet 
at zero field leads to a selection of coplanar states 
below the crossover temperature $T^*/J \approx0.005$
\cite{Chalker92,Zhitomirsky08}. A true long-range
order is not possible in 2D, though 
one can still discuss an asymptotic breaking of the spin-rotation symmetry
at distances smaller than the correlation length $\xi\sim \exp(J/T)$.
The  symmetry breaking in the ensemble of coplanar states
is described by an unconventional octupolar order parameter \cite{Zhitomirsky08}. 
For classical spin models
magnetic octupoles are represented by the on-site third-rank tensor 
\begin{equation}
T_i^{\alpha\beta\gamma} = S_i^{\alpha}S_i^{\beta}S_i^{\gamma} - \frac{1}{5} 
S_i^{\alpha}\delta_{\beta\gamma} - \frac{1}{5}S_i^{\beta}\delta_{\alpha\gamma}  
- \frac{1}{5}S_i^{\gamma}\delta_{\alpha\beta}  \ .
\label{octupole}
\end{equation}
The other common unconventional order parameter is the quadrupolar tensor 
\cite{Andreev84}
\begin{equation}
Q_i^{\alpha\beta} = S_i^{\alpha}S_i^{\beta} - \frac{1}{3}\delta_{\alpha\beta}  \ .
\label{quadrupole}
\end{equation}
Spin tensors (\ref{octupole}) and (\ref{quadrupole}) transform under $l=3$ and 
$l=2$ representations of the $O(3)$ rotation group, respectively. The usual 
(anti)ferromagnetic order parameter ${\bf m}_{\bf q}$ corresponds to the $l=1$ 
representation. In a typical coplanar ground state of the kagome antiferromagnet
spin on a given site may be oriented parallel to any of the three principal directions 
of the $120^\circ$ spin triad. Thus, $\langle S_i\rangle=0$ and the lowest order 
spherical tensor, which captures the $D_3$ (triatic) symmetry of the ensemble of 
coplanar states, is the uniform octupole tensor
\begin{equation}
T^{\alpha\beta\gamma} = \frac{1}{N}\sum_i 
\langle T_i^{\alpha\beta\gamma}  \rangle \ .
\label{octupoleU}
\end{equation}
The quadrupole tensor 
\begin{equation}
Q^{\alpha\beta} = \frac{1}{N}\sum_i \langle Q_i^{\alpha\beta} \rangle
\label{quadrupoleU}
\end{equation} 
is the secondary order parameter and is induced  via the rotationally invariant 
coupling term $Q^{\alpha\beta}T^{\alpha\mu\nu}T^{\beta\mu\nu}$.

In an applied magnetic field the spin symmetry is reduced to the $O(2)$ group. Its irreducible
representations  are labeled by the projection of the angular momentum $l_z$ 
on the field direction. Therefore, different components of the same spin tensor may now
describe different instabilities of the paramagnetic phase:
\numparts
\begin{eqnarray}
l_z = \pm 1 \ , \qquad  Q^{z\alpha}    \ ,  T^{zz\alpha}     \ , \label{lz1}  \\
l_z = \pm 2 \ , \qquad Q^{\alpha\beta} \ ,  T^{z\alpha\beta} \ , \label{lz2}  \\
l_z = \pm 3 \ , \qquad T^{\alpha\beta\gamma} \ ,  \quad  
\alpha,\beta,\gamma= x,y \ .
\end{eqnarray}
\endnumparts
One can arbitrarily select either $Q^{\alpha\beta}$ or $T^{\alpha\beta\gamma}$
to characterize states with $|l_z|=1,2$. In the presence of a uniform 
magnetization $m^z$ the two tensors are coupled via a bilinear term
$m^z Q^{\alpha\beta}T^{z\alpha\beta}$  and become completely equivalent from
the symmetry point of view, see also \cite{Shannon10}.
 
The two coplanar states stabilized in the kagome antiferromagnet at finite fields
break differently the $O(2)$ rotation symmetry. The translationally disordered
Y-state has the residual nematic symmetry $C_2$ and its order parameter is 
given by a $2\times 2$ symmetric traceless matrix (\ref{lz2}).
The $\mathbb{V}$-state may break the translational symmetry, then,
its order parameter is the antiferromagnetic vector of the $\sqrt{3}\times\sqrt{3}$
structure \cite{Zhitomirsky02}. If fluctuations are not able
to select this periodic pattern, the proper order parameter of the 
$\mathbb{V}$-state is transverse vector components of the quadrupolar tensor (\ref{lz1}).
An interesting observation is that the symmetry of the translationally
disordered $\mathbb{V}$-state coincides with the symmetry of 
the transverse ferromagnetic polarization ${\bf m}^\perp$ allowing the interaction
term $m^z Q^{z\alpha} m^\alpha$. For the Heisenberg model the classical
constraint (\ref{constraint}) ensures that ${\bf m}^\perp=0$ so that
the corresponding coupling constant exactly vanishes. 
Nevertheless, the  ground-state constraint may be modified, for example, by
a single-ion anisotropy $D\sum_i S_i^{z2}$.  We have checked that in the 
easy-axis case $D<0$ the anisotropy produces a substantial transverse magnetization
$m^\perp \propto |D|$ in the $\mathbb{V}$-state of the kagome antiferromagnet. 
 
\begin{figure}[t]
\centerline{
\includegraphics[width=0.5\columnwidth]{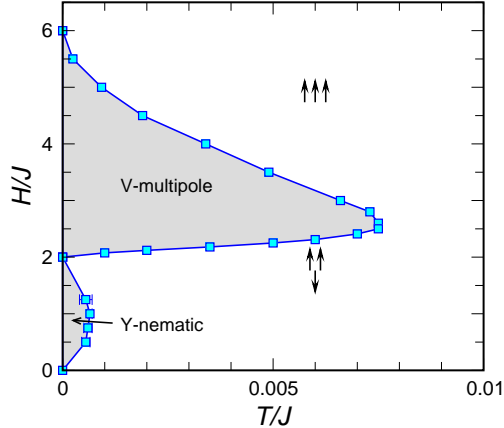}
}
\caption{Magnetic field phase diagram of the Heisenberg kagome-lattice 
antiferromagnet. Transition points determined by the Monte Carlo
simulations are shown by squares. Solid lines are guides for the eye.
}
\label{fig4}
\end{figure}

Partial breaking of the spin-rotation symmetry 
by multipolar order parameters leads, generally, to fractional topological 
defects \cite{Zhitomirsky08}. 
The $C_2$ symmetry of the Y-state is compatible with elementary half-quantum vortices,
whereas  the $\mathbb{V}$-multipolar state supports only usual integer quantized 
defects. The universal jump of the spin stiffness at the KT-transition is 
given by \cite{Chaikin,Korshunov02}
\begin{equation}
\frac{\rho_s}{T} = \frac{2}{\pi \nu^2} \ ,
\end{equation}
where $\nu$ is the vortex winding number. For the half-quantum vortices with 
$\nu=1/2$ the jump in the spin-stiffness amounts to $\rho_s/T  = 8/\pi$.

\subsection{Monte Carlo results}
\label{sec:MCkagome}

\begin{figure}[t]
\centerline{
\includegraphics[width=0.5\columnwidth]{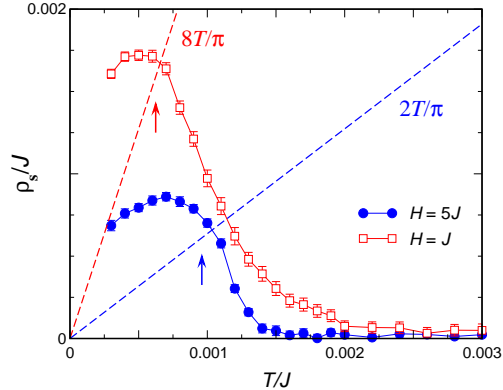}
}
\caption{Spin stiffness of the kagome antiferromagnet 
for two values of the applied magnetic field. 
Points are numerical data for the $L = 36$ cluster.
Vertical arrows indicate positions of the Kosterlitz-Thouless transition
obtained from the susceptibility data:
$T_{\rm KT}/J\approx 6.5\times 10^{-4}$ for $H/J=1$ and
$T_{\rm KT}/J\approx 9.3\times 10^{-4}$ for $H/J=5$.
Dashed lines represent the universal spin stiffness jump (14) for $\nu=1/2$ 
and $\nu=1$.
}
\label{fig5}
\end{figure} 
 
A Monte Carlo investigation of the kagome antiferromagnet in external magnetic field
has been performed using the same simulation protocol as described previously 
for the triangular antiferromagnet in section~\ref{sec:triangular}. 
To obtain a good statistics, one needs, however, to increase the  
number independent cooling runs to 30--50. We have studied
periodic lattices with $N=3L^2$ sites and $L$ in the range 18--96. The obtained phase 
diagram of the kagome antiferromagnet is presented in figure~\ref{fig4}. The 
Kosterlitz-Thouless transition boundaries for the two coplanar states were determined 
from the plots of the rescaled quadrupolar susceptibility with the exact KT-exponent 
$\eta = 1/4$ \cite{Chaikin}. We also measured the spin stiffness of the kagome
antiferromagnet, which is obtained from the previous expression (\ref{rhoS}) 
by applying the renormalization factor $3/4$ \cite{Zhitomirsky08}.
The temperature variation of $\rho_s$ for two values of the external field is shown in 
figure~\ref{fig5}. The stiffness jump at the transition into the Y-state ($H/J=1$) is equal 
with good accuracy to $8T/\pi$. This large jump demonstrates that the corresponding KT-transition 
is mediated by the unbinding of the half-quantum vortices. The jump of $\rho_s$
at $H/J=5$ is substantially smaller and agrees with the standard value $2T/\pi$
expected for the $XY$ vector order parameter.

\begin{figure}[t]
\centerline{
\includegraphics[width=0.5\columnwidth]{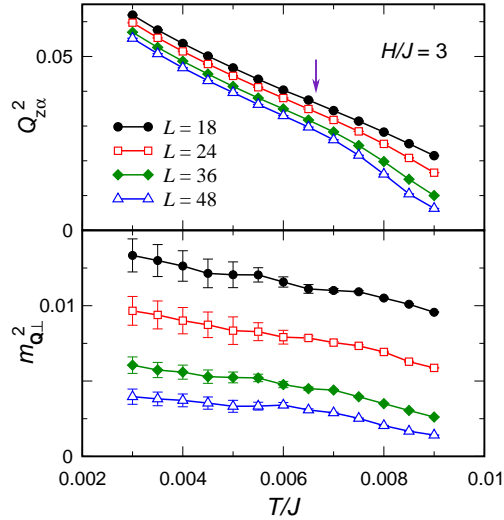}
}
\caption{Temperature dependence of the quadrupole tensor $Q_{z\perp}^2$ (upper panel)
and the transverse antiferromagnetic order parameter of the $\sqrt{3}\times\sqrt{3}$
structure (lower panel) at $H/J=3$. The vertical arrow indicates 
the Kosterlitz-Thouless transition temperature at 
$T_{\rm KT}/J\approx 6.7\times 10^{-3}$.
}
\label{fig6}
\end{figure}

Fractional vortices are incompatible with a conventional antiferromagnetic
ordering, therefore, the spin-stiffness data prove presence 
of the quasi-long-range nematic order in the low-field Y-state.
In order to identify the type of symmetry breaking in the high-field
$\mathbb{V}$-state we consider in figure~\ref{fig6} the temperature evolution 
at $H/J=3$ of  the vector component of the quadrupole tensor 
$Q_{z\perp}^2 = (Q^{zx})^2 + (Q^{zy})^2$ 
and the transverse antiferromagnetic amplitude $m_{\bf Q\perp}$ of 
the $\sqrt{3}\times\sqrt{3}$ configuration. 
The order parameters take on 
a finite value in the presence of a long-range order, but scale down as $1/N$,
if correlations decay exponentially.
The quadrupole tensor $Q_{z\perp}^2$ demonstrates a very weak decrease 
with the lattice size at low temperatures, which is completely consistent with the 
power-law decay $\langle Q_i Q_j\rangle \sim r_{ij}^{-\eta}$ and $T$-dependent 
$\eta <1/4$. The antiferromagnetic order parameter goes down much faster, though 
still slower than $1/N$. This is related to emergence 
of the Coulomb-type correlations that are typical to the constrained spin models,
see, for example, \cite{Henley10}. Overall, the Monte Carlo data show the absence of 
the translational symmetry breaking in the $\mathbb{V}$-state.

Thermal fluctuations have a strong effect in fields around $\frac{1}{3}H_s$, 
where they stabilize the collinear $uud$ configuration between the two coplanar 
phases, see figure~\ref{fig4}. The corresponding 1/3-magnetization plateau
survives to relatively high temperatures $T/J\sim 0.05$--0.2 \cite{Zhitomirsky02}.
The extreme stability of the collinear states is explained by the fact that
they possess the largest possible number of zero-energy modes.
Nevertheless, there is no symmetry breaking related to this selection
and a collinear spin-liquid phase at the 1/3-plateau is connected to the
paramagnetic phase by a simple crossover.

\begin{figure}[t]
\centerline{
\includegraphics[width=0.5\columnwidth]{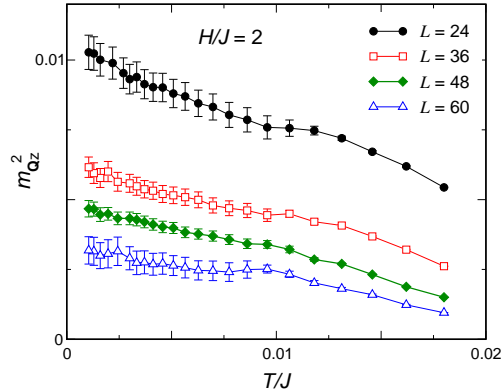}
}
\caption{Temperature variation of the longitudinal antiferromagnetic order parameter of 
the $\sqrt{3}\times\sqrt{3}$ structure  at the 1/3-magnetization plateau, $H/J=2$. 
}
\label{fig7}
\end{figure}

Indeed, the collinear configurations preserve $O(2)$ rotations about the field direction
and the only type of symmetry breaking at the plateau may be related to selection of a specific 
periodic pattern. The harmonic spectra of classical fluctuations are  
identical for different $uud$ structures,
though the anharmonic corrections are, generally, not  \cite{Zhitomirsky02}. 
The two particularly simple spin structures are the $q=0$ configuration
and the $\sqrt{3}\times\sqrt{3}$ state. In the dimer language,
the two states correspond to the columnar array of dimers and the staggered arrangement,
which maximizes the number of three-dimer hexagons.  In our Monte Carlo
simulations we did not find any significant tendency for the $q=0$ ordering
in the classical model, though this configuration is favored by the zero-point
fluctuations in the large-$S$ quantum kagome antiferromagnet \cite{Hassan06}.
The numerical data for the antiferromagnetic order parameter of the longitudinal  
$\sqrt{3}\times\sqrt{3}$ spin structure are presented in figure~\ref{fig7}.
One finds again a fast, but slower than $1/N$ decrease of $m_{\bf Q}^{z2}$
with increasing lattice size. There is no evidence of 
the long-range ordering down to $T/J\sim 5\times10^{-4}$.

The overall behaviour of the antiferromagnetic order parameter is consistent
with the pseudo-dipolar correlations in the 2D Coulomb phase \cite{Henley10}.
At finite temperatures the Coulomb gas description of the ensemble of 
the collinear ground states at $H=\frac{1}{3}H_s$ is extended
by allowing a nonzero density of monomers, which correspond to triangles
with broken constraint condition (\ref{constraint}).  The monomer doping
destroys the power-law correlations at large distances producing a finite 
correlation length \cite{Alet06}. A detailed investigation of the crossover 
from short-distance power-law decay to  exponential decrease at large distances
requires systematic investigation of much bigger lattices.

\section{Summary and discussion}
\label{sec:discussion}

In this paper, we have considered the effect of thermal fluctuations
in external magnetic field for two classical frustrated antiferromagnets 
on 2D triangular and kagome lattices. The common property of the two
models is selection of two coplanar and one collinear spin configurations
by short-range fluctuations. 
For the triangular antiferromagnet such an order by disorder mechanism
produces three ordered magnetic structures, see figure~\ref{fig3}.
The most dramatic effect concerns the collinear up-up-down configuration,
which changes from being marginally stable at the mean-field level  to
having the highest transition temperature among the three states.

For the kagome antiferromagnet, in addition to an arbitrary choice of the spin 
triad subject to the constraint (\ref{constraint})  there is also
a massive degeneracy related to different arrangements of the triad over
the whole lattice. No periodic magnetic structure is selected down
to at least $T/J\sim 4\times 10^{-4}$. We find instead unconventional
magnetic ordering described by multipolar (tensor) order parameters, see 
the phase diagram of the kagome antiferromagnet in figure~\ref{fig4}. 
The most significant change again concerns the $uud$ state,
which does not break any symmetry and corresponds to a collinear
spin liquid phase. Despite producing a pronounced 1/3-magnetization plateau
this state is connected to a  high-temperature paramagnetic
phase by a crossover rather than a  transition.

The surprising feature of the phase diagram of the kagome antiferromagnet,
figure~\ref{fig4}, is an order of magnitude difference between 
transition temperatures of the high-field and the low-field multipolar
state.  Such a remarkable dissimilarity of the two temperature scales 
is produced by a combined effect of (i) 20\% difference in the ground-state 
entropy, see (\ref{N3col}) and (\ref{Ndimer}), and (ii) 
different harmonic spectra with additional zero-energy modes present for
the $\mathbb{V}$-state \cite{Zhitomirsky02}.
One can speculate, therefore, that if other terms are added to
the spin Hamiltonian (\ref{H}), the system would 
more easily find an optimal periodic structure in the $\mathbb{V}$-state
rather than in the more degenerate Y-state. Let us also note,  
that there is an interesting similarity between the phase diagram in figure~\ref{fig4}
and the experimental phase diagram of the 3D hyperkagome antiferromagnet
$\rm Gd_3Ga_5O_{12}$, which orders only in finite magnetic
fields \cite{Schiffer94}. Thus, a Monte Carlo investigation
of the magnetic diagram of the hyperkagome model is a pressing
extension of the present study. 
One should be also intersted in the effect of the single-ion anisotropy on 
the phase diagram of the triangular Heisenberg antiferromagent
in relation to the experimental studies of
$\rm RbFe(MoO_4)_2$ \cite{Svistov03,Svistov06} and 
$\rm Rb_4Mn(MoO_4)_3$ \cite{Ishii09}.

\ack

We acknowledge valuable discussions with V I Marchenko,
R Moessner, Y Motome and K Penc.
Part of this work has been performed within the Advanced Study
Group Program on ``Unconventional Magnetism in High Fields'' at
the Max-Planck Institute for the Physics of Complex Systems (Dresden).

\end{document}